\documentstyle[11pt,aaspp]{article}

\begin{document}

%Remove the following pagestyle command before submission:
%
%Removed \small from \markright argument list so LaTex could run
% file in Socorro - 29 May 95 - Moffett
%\setcounter{page}{0}
%\pagestyle{myheadings}
%\markright{Draft:\protect {Multifreq Obs of the Crab
%Pulsar, {\em Moffett \& Hankins}, Rev: \today } \hfill p\thepage}
%
% Title and abstract page

\title{MULTIFREQUENCY RADIO OBSERVATIONS\\ OF THE CRAB PULSAR}
\author{David A.~Moffett\altaffilmark{1}\altaffilmark{2} and Timothy H.~Hankins\altaffilmark{3}}
\affil{Physics Department, New Mexico Institute of Mining and
Technology\\ Socorro, NM 87801}

\altaffiltext{1}{also National Radio Astronomy Observatory, P.O. Box O,
Socorro, NM  87801} 
\altaffiltext{2}{Electronic mail: dmoffett@nrao.edu}
\altaffiltext{3}{Electronic mail: thankins@nrao.edu}

\begin{abstract}
Previously unseen profile components of the Crab pulsar have been
discovered in a study of the frequency-dependent behavior of its
average pulse profile between 0.33 and 8.4\,GHz.  One new component,
$36^\circ$ ahead of the main pulse at 1.4\,GHz, is not
coincident with the position of the precursor at lower frequencies.
Two additional, flat-spectrum components appear after the interpulse 
between 1.4 and 8.4\,GHz.  The normal interpulse undergoes a
transition in phase and spectrum by disappearing near 2.7\,GHz, and 
reappearing $10^\circ$ earlier in phase at 4.7 and 8.4\,GHz with a new
spectral index.  The radio frequency main pulse disappears 
for $f > 4.9$\,GHz, 
even though it is seen at infrared, optical, and higher energies.
The existence of the additional components at high frequency 
and the strange, 
frequency-dependent behavior is unlike anything seen in other
pulsars, and cannot easily be explained by emission from a simple
dipole field geometry.
\end{abstract}


\section{Introduction}

Since its discovery (\markcite{Staelin \& Reifenstein 1968},
\markcite{Comella et al. 1969}), the Crab Nebula pulsar has been
studied at many wavelengths, including optical, X-ray and
$\gamma$-rays, to understand the properties of the emission mechanism.
Radio emission from the Crab pulsar is unusual because of its giant
pulses, which are very powerful intensity fluctuations varying from
100 to 1000 times the mean intensity that occur at random intervals
from one pulse to several thousand pulse periods.  But the pulsar's
steep radio spectrum and the radio-bright Crab Nebula background make
observations above 1\,GHz difficult with single dish antennas.
Average profiles formed at 4.7\,GHz during the single successful
observation in a decade of attempts
%for over a decade of observations
at the Arecibo 305-m telescope, revealed that the main pulse (MP) to
interpulse (IP) separation appeared to be smaller than at longer
wavelengths, and that additional pulse profile components were present
following the IP.
To investigate this atypical frequency-dependent behavior of the pulse
profile and to overcome the background contribution of the Nebula, the
observing program was continued at the Very Large Array (VLA) of the
National Radio Astronomy Observatory.

\section{Observations}

A series of observations of the Crab pulsar were made between
February 9 and May 27, 1994 at 0.33, 1.4, 4.9, and 8.4\,GHz with the
VLA in its phased-array mode.  The antenna peculiar
phases are determined by observing an extragalactic continuum point
source calibrator and then applied in real time to synthesize a
pencil beam on the sky.  Using this aspect of the interferometer, the
longer spatial wavelengths of the bright Crab Nebula supernova remnant
are resolved. Consequently, when observing the Crab pulsar with the
VLA, the effective system temperature is that of a single 25-meter
dish, but with the collecting area for a point source equivalent to a
130-meter dish.  At the VLA the signal-to-noise ratio of the Crab
pulsar is therefore greater than for the Arecibo telescope by
approximately the inverse ratio of the element areas, or (305\,m /
25\,m)$^2$.

When operating in the VLA's phased-array mode, the received voltages from
each antenna are appropriately delayed, then sampled, summed,
and re-converted to an ``Analog Sum'',
which represents the analog voltage from the synthesized pencil beam. During
our observations, the left- and right-circularly polarized Analog Sum
signals were sent to a MkIII VLBI Video Converter, which can be used
as a dual
14-channel filter bank with bandwidths adjustable in octave steps from
0.125 to 4 MHz. The filter bank is followed by a set of square-law
detectors with adjustable integration time constants and DC
offsets. The channel bandwidths and detector time constants were
selected to optimize the time resolution afforded by dispersion
smearing.  We utilized the Princeton/Dartmouth Mark III Pulsar Timing System
(\markcite{Stinebring {\em et al.} 1992}) to sample and record 14
frequency channels of the left- and right-circularly polarized
signals.  To fold the incoming data at the topocentric period, we utilized the
monthly timing models published by \markcite{Lyne \& Pritchard
(1994)} and the program TEMPO (\markcite{Taylor \& Weisberg 1989}) 
to generate the topocentric
period for each of our observing sessions as input to the timing system.
Each scan was two minutes long, during which an average profile was 
created and saved with a precise time stamp of the scan.
The observed profile consists of 1024 numbers representing the
relative flux density throughout the entire pulsar rotation period.
The timing system clock was referenced to the station standard clock,
and was started by a ten-second tick referenced to UTC through a 
GPS receiver.

Pulsed emission was easily detected at 0.33 and 1.4\,GHz, but emission
at 4.9 and 8.4~GHz was strongly variable and not detected for all the
observing sessions.  Significant emission at 4.9~GHz was detected only
during four of nine sessions of 3-hour duration. Detection of 8.4~GHz
emission occurred only on two sessions.  During the times when the
average profile was seen in a two-minute average, individual ``giant''
pulses could easily be seen on a synchronized oscilloscope
(\markcite{Hankins \& Moffett 1996}).  The typical interval over
which the pulsar could be seen above the noise level at 8.4\,GHz
was 10 to 30 minutes, and these did not often
repeat in a single observing session.  Unfortunately, we could not
observe separate frequencies simultaneously, and so we cannot make
quantitative estimates as to how long the emission was absent during a
session, since our procedure was to search at other frequencies if
detectable emission could not be found after 10 to 20 minutes.  On one
occasion at 8.4~GHz, pulses occurred in bursts where a pulse was seen
in nearly every period for up to a minute.  We tentatively attribute
this effect to diffractive interstellar scintillation, which is
usually associated with short time-scale intensity fluctuations.

\section{Timing Analysis}

Total intensity profiles were formed by adding two minute average
profiles from each filter bank channel after accounting for the
appropriate dispersion measure delay between channels.  Each profile
thus summed was cross-correlated with a high signal-to-noise standard
profile, or {\em template}, to determine the phase and ultimately the
time of arrival of the pulsar waveform with respect to a reference or
{\em fiducial} point on the template.  At 0.33 and 1.4~GHz, the
signal-to-noise ratio was high enough that the respective templates
could be formed from stacked scans.  At 4.9 and 8.4~GHz, the
signal-to-noise was low, so a Gaussian template was used for these
frequencies.  Since the waveform of the Crab is frequency dependent,
the templates were aligned to the centroid of the main pulse.  However
the main pulse was not a suitable fiducial point at 8.4\,GHz, and so
its alignment with the other frequencies was constrained by the timing
model and the dispersion measure.  Because of the instrumental
smoothing response of the detectors, the position of the MP at 0.33
and 1.4\,GHz was determined after deconvolving a model of the 
detector response function.

After observing, we used the same input timing models of
\markcite{Lyne \& Pritchard (1994)} to compare with the actual times
of arrival (TOA) from our scans.  The standard deviation of our timing
residuals for any single session was on the order of 4\,$\mu$s at 1.4
GHz, which is considerably smaller than the standard deviation of the
residuals from the average time of arrival from each day. The
residuals from 4.9 and 8.4 GHz were higher, from 15 to
40\,$\mu$s.  Since the signal-to-noise was low for the higher
frequencies, we attribute the scatter of 4.9 and 8.4 GHz residuals
from a single session to estimation error from radiometer noise.
Shifts of 50 to 100\,$\mu$s in the 1.4\,GHz arrival times were
found to occur from one session to the next.  Such jumps have been
seen in optical and radio timing observations of the Crab
(\markcite{Boynton et al. 1972}, \markcite{Groth 1975},
\markcite{Lyne, Pritchard, \& Smith 1988}), and have been described
as ``timing noise''.  In a study of the timing
noise phenomenon, \markcite{Cordes (1980)} found the process to 
be a random walk
process on short time-scales; but over longer periods of time,
\markcite{Lyne, Pritchard, \& Smith (1988)} found that the deviation
in arrival times has a quasi-sinusoidal component with a 20-month
period whose amplitude varies from 5 to 10 milliseconds.  
Over our short observing term, 
our arrival time deviations would show no evidence of this
periodic component, and the arrival times from one observing session
to the next would appear to be random.

Solutions for dispersion measure were made from individual
multifrequency observation sessions.  During a session from February
22 to 23, three frequencies were observed and the 
dispersion measure (DM) found during this
session was $56.826 \pm 0.010$ pc-cm$^{-3}$, which agrees with Lyne
\& Pritchard's average value of $56.827 \pm 0.005$ pc-cm$^{-3}$ for
February 1995.  Other determinations were made from
observations at 1.4 and 4.9\,GHz on February 14 and 19, and May 14.
No significant departures of DM from Lyne \& Pritchard's models for
February and May were found beyond three times the estimated error of
0.01 pc-cm$^{-3}$, and there is no evidence for superdispersion delay.  After
removing the time delay for a fixed DM, the profiles from 0.33, 1.4,
and 4.9\,GHz were found to align at the position of the main pulse,
which was chosen earlier as the reference point for determining the
arrival time of the averaged pulses.  The 8.4\,GHz profile was found
to align with other profile components seen at 4.9\,GHz.

\section{Multifrequency Properties}

\begin{figure}
\epsfysize=7.5truein
\epsfbox{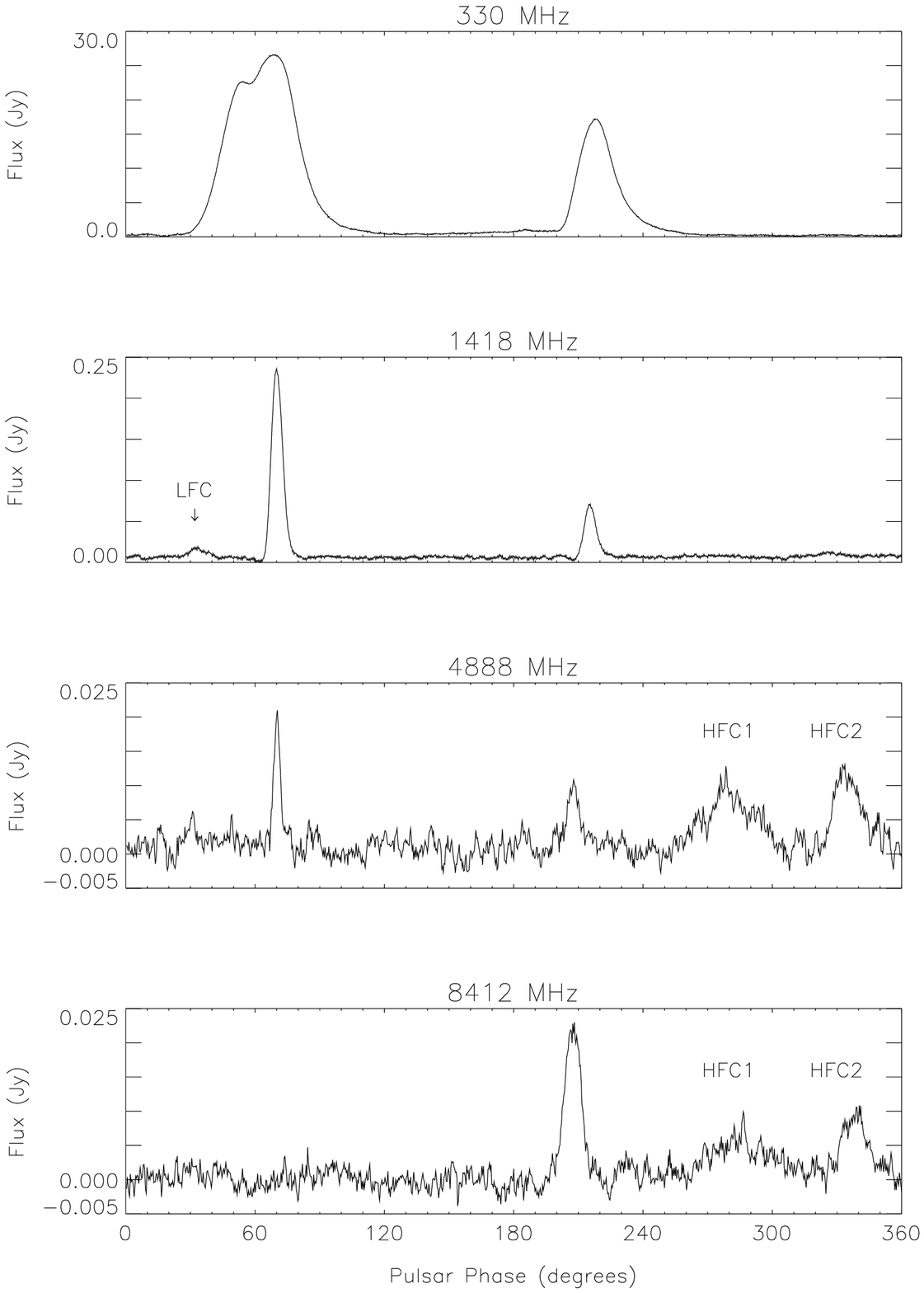}
\caption{Aligned VLA profiles of the Crab pulsar.\label{f1}}
\end{figure}

Figure \ref{f1} shows the aligned, average profiles recorded
at 0.33, 1.4, 4.9, and 8.4\,GHz with the phased VLA. 
The amplitude scale at each frequency
was estimated from the off-pulse radiometer noise, antenna and
receiver parameters, integration time, and expected contribution from
the Crab Nebula.

In the 0.33\,GHz profile one can see the precursor, MP and IP.  The
precursor and MP overlap, due to a dispersion broadening of 1.6\,ms
across the 0.125\,MHz bandwidth filters and a 1-ms detector time
constant.  One feature to note is a broad non-zero component or {\em
emission bridge} between the MP and IP, which has been detected
previously by \markcite{Rankin {\it et al.} (1970)} at 196.5\,MHz, 
by \markcite{Manchester,
Huguenin, \& Taylor (1972)} at frequencies from 114 to 159\,MHz, and 
by \markcite{Vandenberg {\it et al.} (1973)} at 111.5 and 196\,MHz.  The
estimated flux density of the average profile is 0.88 Jy,
nearly the same as estimated from the spectrum compiled 
by \markcite{Sieber (1973)}.

At 1.4~GHz, the precursor has vanished due to its steep spectral
index, leaving only the MP, IP, and a weak but distinct low-frequency
component $\sim 36^\circ$ ahead of the MP (which we will refer to as
LFC).  The appearance of this component, distinct from
the precursor, has been verified by observations at 608 and 1400\,MHz
\markcite{(Lyne \& Pritchard 1995)}, and can be seen as a slight rise
above the noise level at 4.8\,GHz. This new component may be present
at 0.33\,GHz, but better time resolution would be required to separate
it from the precursor.  The estimated integrated flux density of this
profile is 6.3 mJy; only half of the expected value.

\begin{figure}
\epsfysize=7.5truein
\epsfbox{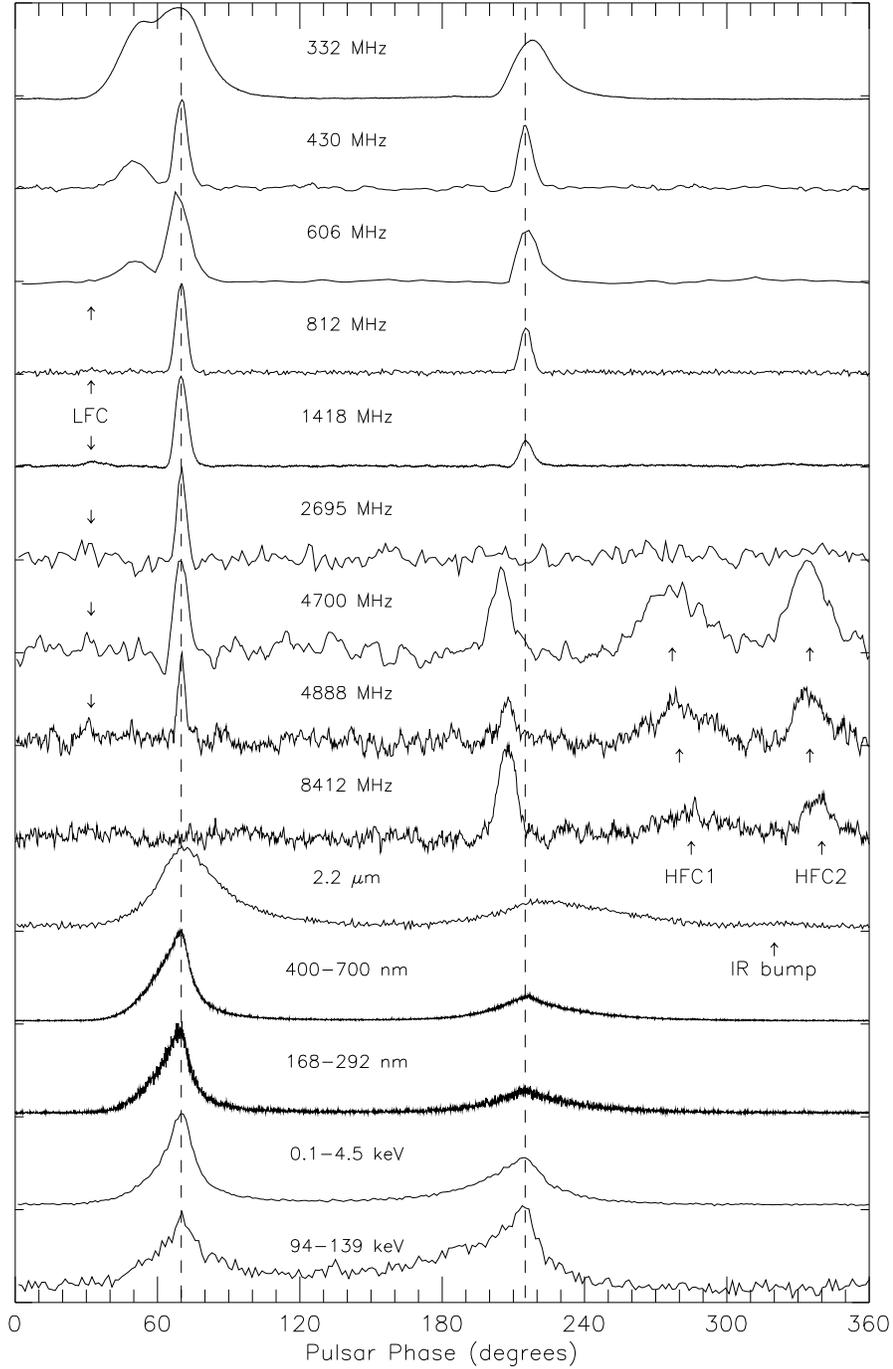}
\caption{A summary of Crab pulsar profiles.\label{f2}
332, 430, 606, 1418, 2695, 4700, 4888 and 8412\,MHz: this work; 
812\,MHz, 2.2$\mu$m: Lundgren, Cordes, \& Beckwith (1995), 
400-700 and 168-292 nm: Percival et al. (1993); 0.1-4.5 keV: 
Harnden \& Seward (1984), 94-139 keV: Ulmer et al. (1994).}
\end{figure}

At 4.9 GHz, the profile becomes quite complex.  The profile clearly
shows four components despite the lower signal-to-noise ratio.  Since
our profiles are aligned, we identify the first two narrow components
as the MP and IP.  The two broader components (which we will refer to
as HFC1 and HFC2) following the IP have similar pulse energies at 4.9
and 8.4\,GHz, indicating that they are nearly flat-spectrum
components.  They are coincident in pulse phase with the weak features
appearing in the 1.4\,GHz profile.  Another striking feature of this
profile is that {\it the IP appears to shift earlier in phase}.  Both
the shift and the existence of additional components confirm the
4.7\,GHz profile recorded in 1981 at the Arecibo telescope (see Figure
\ref{f2}).  The integrated flux from our estimate is 0.47 mJy,
only slightly higher than expected (0.35 mJy).

In the 8.4\,GHz profile, the main
pulse disappears altogether.  The estimated spectrum from 0.33 to
4.9\,GHz suggests that the main pulse has dropped below the noise
level.  However, the apparent interpulse seems to have 
become more energetic from
4.9 to 8.4\,GHz.  Our estimate of the flux density at 8.4\,GHz is 0.61 mJy,
10 times that extrapolated from lower frequencies.  An
independent continuum observation at 8.4\,GHz by Frail (1995) has
placed the flux density of the pulsar at $0.5 \pm 0.1$ mJy.  Thus the
spectrum of the pulsar seems to become flatter at high frequencies.

In Figure \ref{f2} we
have plotted the VLA profiles in a summary of normalized waveforms 
from other radio frequencies and energy bands.  Profiles at 0.43, 0.6,
2.7 and 4.7\,GHz were obtained from Arecibo and Effelsberg archival
data recorded by Hankins.  Though the VLA profiles are aligned after
removing dispersion delay, all other profiles in the figure have been
arbitrarily aligned by the peak of the main pulse only.  We note, that
radio and higher energy waveforms observed by other authors have been
shown to align absolutely as in the figure.

By comparing all these profiles, we can identify even more properties.
The LFC can be seen from 0.6 to 4.8\,GHz, just
barely above the noise level at some frequencies.
At 2.7\,GHz,
the interpulse disappears, then it reappears at 4.7\,GHz, 
$\sim 10^\circ$ ahead of its low frequency position.
Then between 8.4\,GHz and $2.2\,\mu$m, the profile evolves from having
only three components, with the main pulse missing, back to two
components, with much broader MP and IP.
Though the MP and IP dominate emission in the infrared, we note the
existence of a small rise of emission after the interpulse, in the
same region of phase as radio components HFC1 and HFC2.  
%This low
%level emission may correspond to an extra component seen in the IR
%profiles recorded by Becklin {\it et al.} (1973).  
At higher energies,
the shape of the MP and IP remains virtually constant, with a bridge
of emission between the two, similar to the radio emission bridge seen
at 0.33\,GHz.
  
Gaussian waveforms were fitted to the relative positions of the
profile components in order to find the width and phase separation of
components with frequency.  The
half-power widths of the precursor and HFC2 are comparable, while the
width of the interpulse was found to increase monotonically with 
frequency.
The decrease of $\Delta\phi_{\rm MP-IP}$ is suggestive of a 
radius-to-frequency mapping phenomenon that has
been seen in profiles of conal pulsars (Cordes 1978, Hankins \&
Fowler 1986, Thorsett 1991), but the discontinuous jump of $\Delta\phi_{\rm MP-IP}$ 
from 1.4 to 4.7\,GHz is not consistent with the smooth change with frequency 
in component separation found for conal pulsars.

\section{Discussion}

The appearance of additional profile components at high radio frequencies
(and possibly IR) and the change of profile morphology between radio
and higher energies further complicates the problem of where the Crab
pulsar's emission originates and how it is generated.  
The average profiles presented here are not typical of other pulsar 
profiles; there are up to six 
different regions in pulse phase where emission occurs.  
Of the millisecond pulsars, one is known to
have more components, PSR\,J0437-4715, with up to 8 components or more 
(\markcite{Manchester \& Johnston 1995}).  
Though not rotating as fast, the Crab's
velocity-of-light cylinder is still relatively close to the surface, where 
multipole fields may exist \markcite{(Kuz'min 1992)}.
The HFC components seen at 4.7, 4.9 and 8.4\,GHz, and the
third component at $2.2\,\mu$m in Figure \ref{f2} may stem from
emission originating in multipole fields near the star's surface.

The discovery at 1.4\,GHz of the LFC component leading the MP at
nearly twice the phase separation of the MP and precursor might be
evidence for a core/cone set.  The half-power widths of the precursor
at 0.33, 0.43, and 0.6\,GHz are the same as core component widths
predicted from the empirical core width--period relationship established
by Rankin (1993) for a near orthogonal rotator.
Also, the LFC-to-MP separation is almost the same as the inner conal
diameter predicted by Rankin for a pulsar of this period.  
The steeper spectrum of the precursor relative to the LFC also supports
their interpretation as core and cone components, respectively.
But then if we interpret the LFC--MP pair as conal components 
(with similar spectral indices), we must rationalize their dramatically 
unsymmetric amplitudes.  On the other, if the MP is not one of a conal
component pair, but originates in the
outer magnetosphere, then the lagging conal component mate of the LFC 
may just be buried under the MP.

The disappearance of the MP at 8.4\,GHz should not surprise us.
If the spectral index remains constant between 0.33 and 4.9\,GHz, then the MP
should lie below the noise level at 8.4\,GHz.  We have seen giant
pulses from the MP at this frequency, but when averaged in time,
no component appears. 
For it to reappear at higher energies, both the beam and the emission
mechanism must make a significant change.  
The IP does undergo a transition within the range of radio
frequencies observed here.  The spectral index from
0.33 to 1.4\,GHz suggests that it should be spectrally dead by
4.7\,GHz.  The lack of an IP at 2.7\,GHz tends to support this 
despite the profile's modest signal-to-noise.  The apparent 
IP at $\nu \geq $ 4.7\,GHz
could very well be a {\em new component}, much like the other HFCs, with a
different spectral index from that of the low-frequency IP.  

The high frequency components, HFC1 and HFC2, appear from
1.4\,GHz to 8.4\,GHz, and perhaps contribute to emission
seen in the IR at $2.2\mu$m.  They have a flatter spectrum than the MP
and IP, and at high frequency they dominate the profile, tending  
to flatten the spectral index of the pulsar.
This property has been seen in other pulsars at high frequency
(\markcite{Wielebinski {\it et al.} 1993}) and for the Crab this
should hold true
-- because the energy spectrum between the radio and IR band has a
large discontinuity which could only be accounted for by a turn-up of
the spectrum (see figure 4--2, \markcite{Manchester \& Taylor 1977}).  

We can make no interpretation of the magnetic field
geometry from the pulse morphology without full polarization information.
Weak linear polarization in both the MP and IP was found in observations
by Manchester, Huguenin, \& Taylor (1972) between 110 and 160\,MHz and 
by Manchester (1971) at 430 and 1400\,MHz.  And the precursor is well
known to be strongly polarized.  However, no significant
sweep in polarization position angle was found across the precursor, 
MP and IP.
If the rotating vector model of Radhakrishnan and Cooke (1969) holds
true, then the lack of position angle variation
suggests that emission may be from one pole as it sweeps across
the line of sight.  However the optical polarization position angle
sweep (Kristian {\it et al.} 1970) suggests that emission comes from 
two poles (Narayan \& Vivekanand 1982).  A new model and interpretation
by \markcite{Romani \& Yadigaroglu (1995)} 
suggest that optical and other high energy 
emission come from the last open field
lines above one pole in an outer potential gap (Cheng, Ho, \& Ruderman
1986) and exhibits the same observed polarization sweep.
Since the radio MP and IP align with the high energy components,
their radio emission also comes from the outer gap above one pole, and the highly polarized precursor originates above the 
polar cap on the opposite side of the pulsar.

\begin{figure}
\epsfbox{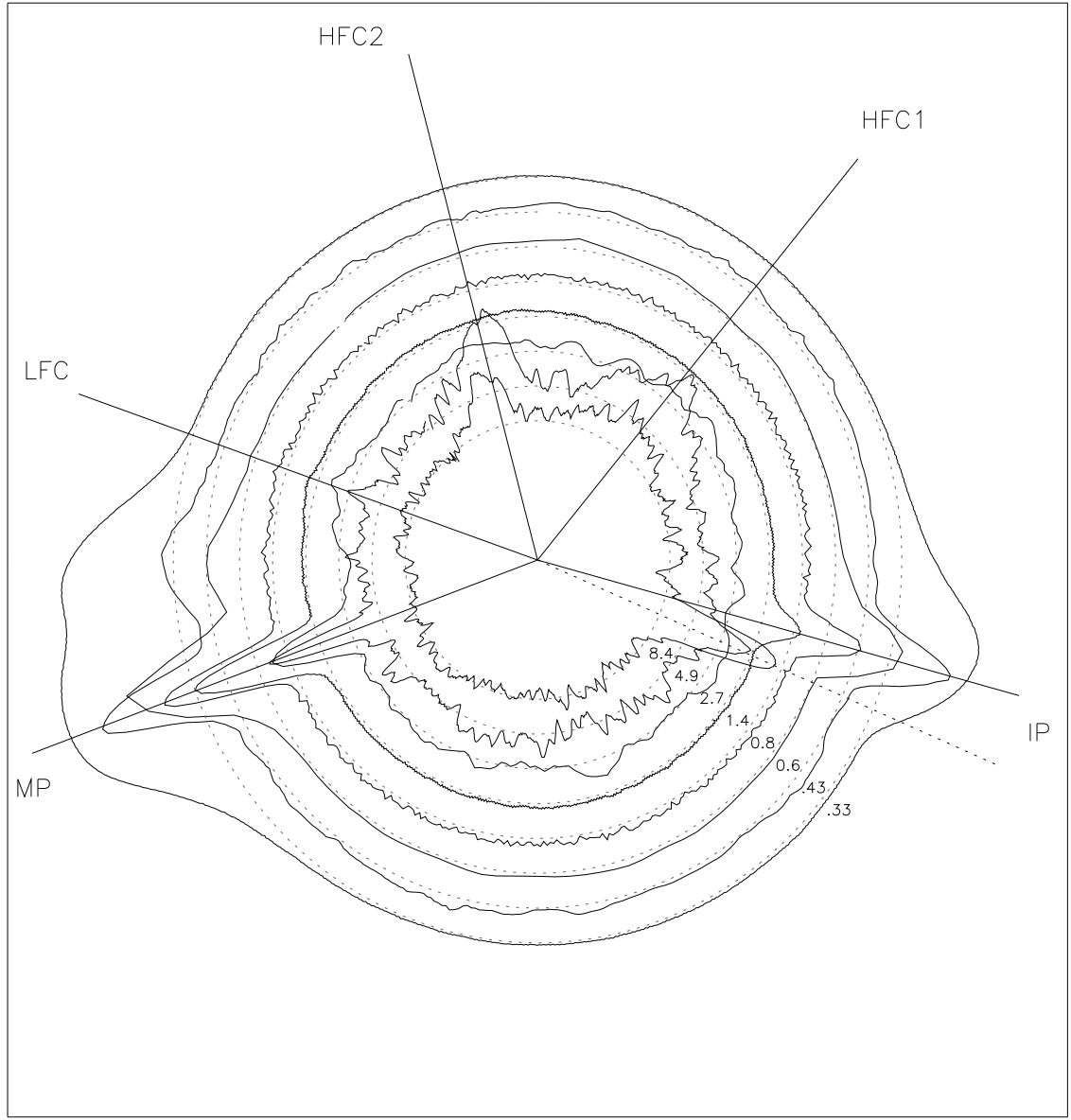}
\caption{A polar plot of Crab pulsar profiles at radio frequencies 
shown in Figure 2.  Frequencies in radial order:
8412, 4888, 4700, 2695, 1418, 812, 606, 430 and 332\,MHz. \label{f3}}
\end{figure}

The profiles do exhibit some apparent symmetry in the placement of
components.  In Figure \ref{f3}, we have arranged average profiles
in a polar plot from 8.4\,GHz in the inner circle to 0.33\,GHz in the
outer circle.  One can easily see the phase shift of the IP from 1.4
to 4.7\,GHz, and the near $180^\circ$ separation between the low
frequency IP and the line denoting the phase of the LFC.  However, the
pulse profile does not appear to be symmetric about this line.  The
only symmetrical features we can derive from this plot are the MP and
IP pair, and the HFC1 and HFC2 pair.  Though the bisectors of these
two pairs of components are less than $180^\circ$ apart, there seems
to be a physical symmetry.  At low frequency, the MP and IP have
nearly the same phase separation and spectral index; a common feature
shared by the HFC components at high frequency.  If the MP and IP
originate in the outer magnetosphere as suggested by \markcite{Romani
\& Yadigaroglu}, then the HFC components may also originate in a
similar region, but on a separate emitting surface with a smaller
opening angle.

\section{Conclusion}

In a multifrequency study of the Crab pulsar, we have found new and
unusual components that defy explanation by emission from a simple
dipole field geometry.  Two of the new components have
a flatter spectrum than the main pulse and interpulse, and so they
survive to become visible at high radio frequencies, possibly even at
infrared energies.  The interpulse is replaced between 2.7 and 4.7\,GHz with a
new component exhibiting a much different spectrum.  And a new
component found $36^\circ$ ahead of the main pulse may be a leading
conal component to the core emission of the precursor.
We are making polarimetric observations at frequencies higher 
than 1\,GHz of the the new components to understand their emission
and location in the Crab's magnetosphere.  We find
that the occasional ``giant'' pulses from the Crab are emitted mostly
at the phase of the MP and less often at the IP, but never at the
phases of the other components.  Since there seem to be no ``giant''
pulses at optical or $\gamma$-ray wavelengths where the emission is
incoherent (\markcite{Lundgren 1994}), the giant pulse phenomenon must
be related to the degree of radio emission coherence, rather than energetic
particle production. The emission region conditions must, therefore be
similar at the MP and IP, but the LFC, HFCs, and precursor must be
significantly different.

The VLA is part of the National Radio Astronomy Observatory, which is
operated by Associated Universities Inc., under 
a cooperative agreement with the National Science Foundation.
The Arecibo Observatory is part of the National
Astronomy and Ionosphere Center, which is operated by Cornell
University under contract with the National Science Foundation. This
work has been conducted with partial support of NSF grant
AST-9315285. DAM acknowledges the support of NRAO as a
Junior Research Fellow, and also thanks D.~Nice of NRAO for software
support and advice during the data reduction process of this project.
We thank J.~Rankin for valuable background
information and discussion of the results, and we thank J.~Gil for
a provocative review.

\vfill\eject

\vfill\eject

\end{document}